\documentclass[pra,aps,twocolumn,showpacs,superscriptaddress,floatfix,amsmath,amssymb,nofootinbib]{revtex4}
\usepackage{amsfonts}
\usepackage{amsmath}
\usepackage{amssymb}
\usepackage{graphicx}
\usepackage{braket}
\usepackage{enumitem}
\usepackage{xcolor}
\usepackage{xifthen}
\usepackage{hyperref}

\newcommand{\beq}{\begin{equation}}
\newcommand{\eeq}{\end{equation}}

\newcommand{\beqa}{\begin{eqnarray}}
\newcommand{\eeqa}{\end{eqnarray}}
\newcommand*{\Ca}{\texorpdfstring{\text{$^{40}$Ca$^+$}}{Ca40}}

\begin{document}

\title{Multi-GHz repetition rate, multi-watt average power, ultraviolet laser pulses for fast trapped-ion entanglement operations}
%alternative titles:
% \title{Ultraviolet, Multi-GHz trapped-ions entangling gate source with average power over 1 W}
% \title{Generation of 5 GHz ultraviolet laser pulse bursts with average power of 5 W for a fast trapped-ion entanglement operation}
% \title{Generation of 5 W average power pulse bursts in the ultraviolet at multi-gigahertz repetition rates for a fast trapped-ion entanglement operation}
\author{M. I. Hussain}
\email{mahmood.hussain@uibk.ac.at}
\author{D. Heinrich}
\author{M. Guevara-Bertsch}
\affiliation{Institute for Quantum Optics and Quantum Information (IQOQI), Austrian Academy of Sciences, Innsbruck 6020, Austria}
\affiliation{Institute for Experimental Physics, University of Innsbruck 6020, Austria}
\author{E. Torrontegui}
\affiliation{Instituto de F\'{\i}sica Fundamental IFF-CSIC, Calle Serrano 113b, 28006 Madrid, Spain}
\affiliation{Departamento de F\'{\i}sica, Universidad Carlos III de Madrid, Avda. de la Universidad 30, 28911 Legan\'es (Madrid), Spain}
\author{J. J. Garc{\'\i}a-Ripoll}
\affiliation{Instituto de F\'{\i}sica Fundamental IFF-CSIC, Calle Serrano 113b, 28006 Madrid, Spain}
\author{C. F. Roos}
\author{R. Blatt}
\affiliation{Institute for Quantum Optics and Quantum Information (IQOQI), Austrian Academy of Sciences, Innsbruck 6020, Austria}
\affiliation{Institute for Experimental Physics, University of Innsbruck 6020, Austria}

\begin{abstract}
The conventional approach to perform two-qubit gate operations in trapped ions relies on exciting the ions on motional sidebands with laser light, which is an inherently slow process.
%relies on the excitmotional-mode frequency, %which is slow.
One way to implement a fast entangling gate protocol requires a suitable pulsed laser to increase the gate speed by orders of magnitude.
However, the realization of such a fast entangling gate operation presents a big technical challenge, as such the required laser source is not available off-the-shelf.
For this, we have engineered an ultrafast entangling gate source based on a frequency comb.
The source generates bursts of several hundred mode-locked pulses with pulse energy $\sim$800 pJ at 5 GHz repetition rate at 393.3 nm
and complies with all requirements for implementing a fast two-qubit gate operation. Using a single, chirped ultraviolet pulse, we demonstrate a rapid adiabatic passage in a Ca$^+$ ion. 
To verify the applicability and projected performance of the laser system for inducing entangling gates we run simulations based on our source parameters. The gate time can be faster than a trap period with an error approaching $10^{-4}$.

\end{abstract}  	
%{03.75.Kk,05.60.Gg,37.10.Gh}
\maketitle

\section{Introduction}

Trapped atomic ions are a well-known resource for testing and implementing quantum information and quantum computation protocols \cite{schmidt2003realization,gulde2003implementation,haffner2008quantum,harty2014high,monz2016realization}.
One of the challenges to execute complex quantum algorithms is to increase the number of gate operations that can be carried out within the coherence time of the ion.
Given a fixed coherence time, developing faster gate operations is a promising strategic approach to address the aforementioned problem.
Well-established and routinely used entangling gate schemes depend on the Coulomb-coupled normal-modes of motion of the ions \cite{cirac1995quantum,molmer1999multiparticle}.
The typical time-scale of the motional modes of the ion is on the order of microseconds, posing a restriction on the speed of the entangling gate operation.
Therefore, implementing faster entangling gate schemes will be a major advancement towards scalable quantum computing.
In this pursuit, a scheme was proposed by Garc{\'{i}}a-Ripoll \cite{garcia2003speed} which exploits state-dependent momentum kicks rather than spectrally resolved motional sidebands to realize a two-qubit entangling gate operation faster than the motional mode period.

The idea is to expose the ion to counter-propagating resonant laser pulses, where each pulse provides a state-dependent momentum kick on the ion by coherent population transfer between the ground and an excited state.
The momentum kicks drive the motional modes of the ions and, given precise control over the timing of incident pulses, force the modes to follow
a closed trajectory in phase space which creates a relative phase between qubit states to generate entanglement \cite{garcia2003speed}.
%The fast gate scheme exhibits multiple motional sideband excitation, hence favours making the gate operation faster than the trap period.
The fast gate scheme does not rely on resolving motional sidebands and hence can be completed in less than a trap period.
%has an area of $\int_0^{\delta t}\Omega(\tau)d\tau=\pi$ $\pi$. Every $\pi$-pulse. 
Attempts are being made to realize fast gate operations by a number of research groups \cite{campbell2010ultrafast,mizrahi2013ultrafast,schafer2018fast,zhang2020submicrosecond}.
Rydberg-mediated entangling gate operations have been proven as a route towards faster gate operations: a two-qubit gate time of 700 ns is achieved to generate a Bell state with 78\% fidelity \cite{zhang2020submicrosecond}. A recent study reported the use of  amplitude-modulated pulses for producing a two-qubit gate with a gate time of 480 ns \cite{schafer2018fast}.
Non-resonant Raman ultraviolet (UV) pulses were used to excite a single trapped-ion \cite{campbell2010ultrafast,mizrahi2013ultrafast} and a proof-of-principle demonstration of a two-qubit phase gate with a gate time greater than the trap period has been reported outside the Lamb-Dicke regime, i.e., the gate operation is insensitive to the thermal motional state of the ions \cite{Wong-Campos2017}.

\begin{figure*}
\includegraphics[width=15cm,height=9cm,keepaspectratio]{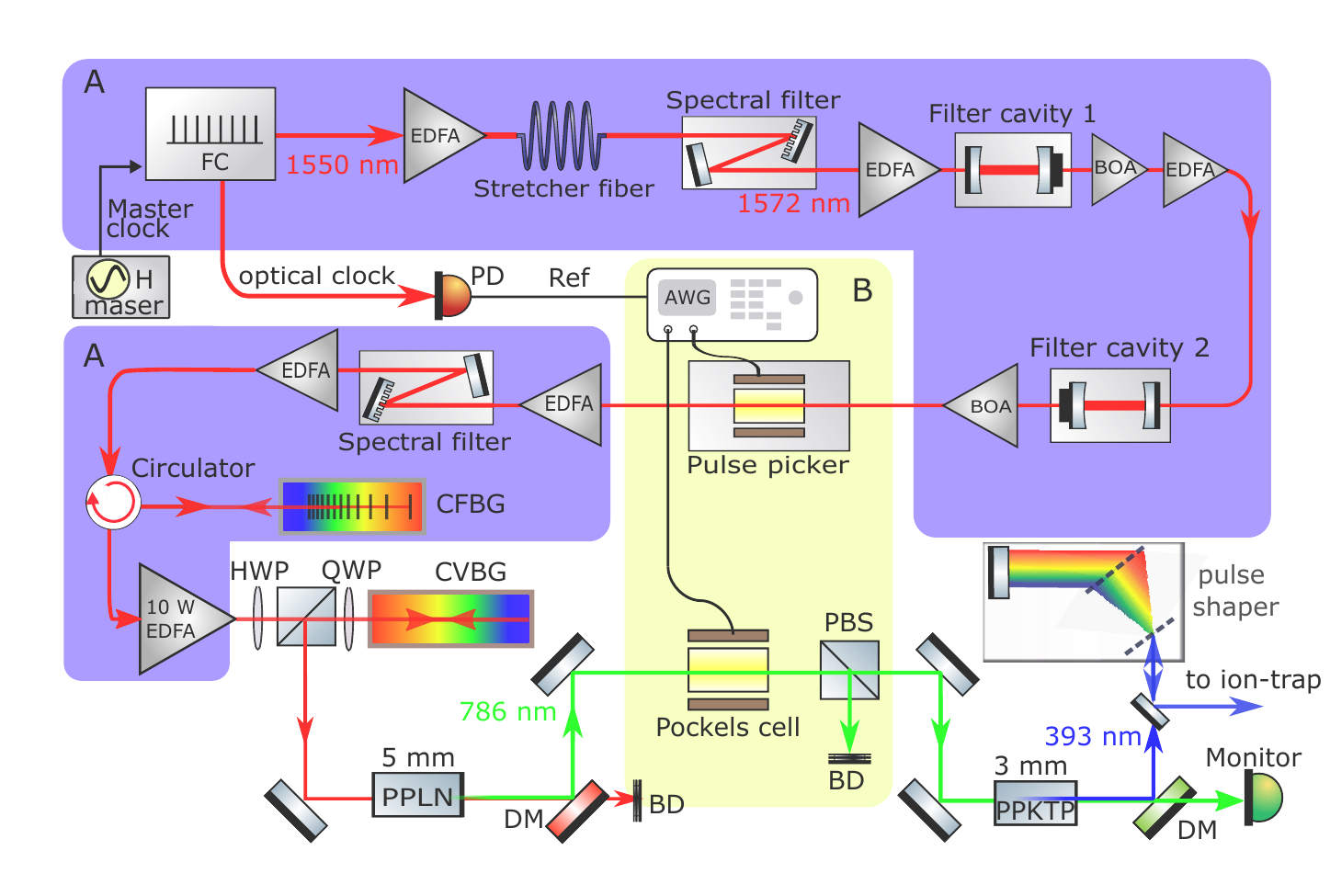}
\caption{Schematic of the ultrafast pulsed laser source. (A) The colour panels (purple) represent mostly the in-fiber part of the laser system.
In these sections, chirped laser pulses at 5 GHz repetition rate are created  with the right optical spectrum to reach the desired wavelength for high power amplification.
Subsequently, amplified pulses travel through free space where nearly transform-limited pulse compression is achieved in CVBG and frequency up-conversion takes place via two non-linear crystals. The filtered UV pulses coming out of PPKTP crystal are steered into the pulse shaper before sending them to the ion trap via optical fiber.
(B) The yellow panel depicts our pulse-switching scheme in infrared and visible wavelengths. Black arrows/lines correspond to electronic signals.
AWG: arbitrary waveform generator, BD: beam dump, 
BOA: booster optical amplifier,
CFBG: chirped fiber Bragg grating,
CVBG: chirped volume Bragg grating, DM:
dichroic mirror, EDFA: erbium-doped fiber amplifier, FC: optical frequency comb, 
HWP: half-wave plate,
PBS: polarizing beam splitter, PD: photo
detector, PPLN: periodically poled lithium niobate, PPKTP: periodically poled potassium titanyl phosphate, 
QWP: quarter-wave plate,
H maser: passive hydrogen maser.}
%{\itshape b)} Corresponding phase as a function of $N$.}
\label{setup}
%\end{center}
\end{figure*}
There are ongoing efforts to develop high average power and high repetition rate, ultrafast UV and extreme-UV laser sources \cite{samanta2014yb,emaury2015compact,wang2015bright,chaitanya2015high,kottig2017generation} for a range of applications and particularly for quantum computation to expedite entangling gates in trapped ions \cite{hussain2016ultrafast}.
In the latter, the difficulty lies in achieving a higher repetition rate ($\gg$ 1/trap period) and sufficiently high pulse energy to enable state-dependent kicks, while restricting the wavelength range to a specific narrow-band ultraviolet spectrum. 

In the present work, we generate frequency-quadrupled UV, picosecond laser pulses for fast entangling gate operations in trapped \Ca{} ions.
The center frequency of the laser source resonates with a strong atomic dipole transition (393 nm) in \Ca{}.
The pulse duration ($\sim$1 ps) is much shorter than the excited state lifetime (6.9 ns) of the ion in order to suppress spontaneous emission.
The combination of a high repetition rate of 5 GHz and a fast pulse picker provides precise control over the timing of picked pulses relative to the ion motion.
At 5 GHz repetition rate, bursts of a few hundred of UV laser pulses are generated with an estimated average power of 5 W.
At a reduced pulse repetition rate of 1.25 GHz, we estimate an average power of over 1 W.   
In both cases, the corresponding pulse energy is roughly 10 times more than previously utilized for a coherent population transfer with a probability of over 90\% \cite{Heinrich2019}. This enhanced pulse energy will enable us to split laser pulses for inducing counter-propagating kicks on the ions.

%Therefore, pulse energy remains remains same as at 1.25 GHz. 
%

The paper is structured as follows:
In section \ref{seed} we briefly describe the fundamental oscillator and related components that seed the laser source.
Section \ref{switch} explains the single pulse switching of 5 GHz fast laser pulses to create pulse sequences for the implementation of a gate.
The methods and generation of UV pulses is thoroughly discussed in section \ref{UV}.
Numerical simulations have been carried out based on our source parameters detailed in section \ref{simulation}. Section~\ref{sec:TrappedIonExperiment} presents data demonstrating a rapid adiabatic passage in a single ion subjected to a single UV pulse.
Finally, we conclude the paper in section \ref{conclude}.

%
%
%
%\section{Laser source}
\section{5 GH\texorpdfstring{\MakeLowercase{z}}{z} telecom band frequency comb}
\label{seed}
The seed oscillator is based on a fiber laser frequency comb produced by Menlo Systems. The mode-locked erbium-doped fiber cavity produces 75 fs laser pulses at a center wavelength of 1550 nm with a fundamental repetition rate of 250 MHz.
A passive hydrogen maser serves as a master clock with a fractional frequency instability of $\leq$9 $\times$ 10$^{-15}$/hr to lock the carrier-envelope offset and laser pulse repetition rate.
Multiple erbium-doped fiber amplifiers (EDFAs) and booster optical amplifiers (BOAs) have been installed to compensate for insertion losses at various points in the laser set-up.
Laser pulses out of the seed oscillator are amplified in the first pre-EDFA and chirped by the stretcher fiber as shown in panel A of Fig. \ref{setup}.
Chirped pulses travel through a spectral filter
where the spectral bandwidth is reduced to 8 nm with a new center wavelength of 1572 nm (4$\times$393 nm).
The repetition rate is ramped up to 5~GHz by a filter cavity with a free spectral range such that it transmits only spectral modes spaced 5 GHz apart and suppresses all others.
The same process is repeated in the second filter cavity to increase the extinction ratio between transmitted and suppressed comb modes.
We integrate a BOA after the second filter cavity to reduce the pulse-to-pulse amplitude noise by saturating the amplifier \cite{mccoy2004intensity}.
Both cavities are locked to an auxiliary continuous-wave laser which in turn is locked
to a mode of the frequency comb as detailed in reference \cite{Heinrich}.
%%%
%%%
%%%
%%%
%%%
\section{Pulse patterning}\label{switch}

\begin{figure}
\includegraphics[width=8cm,height=9cm,keepaspectratio]{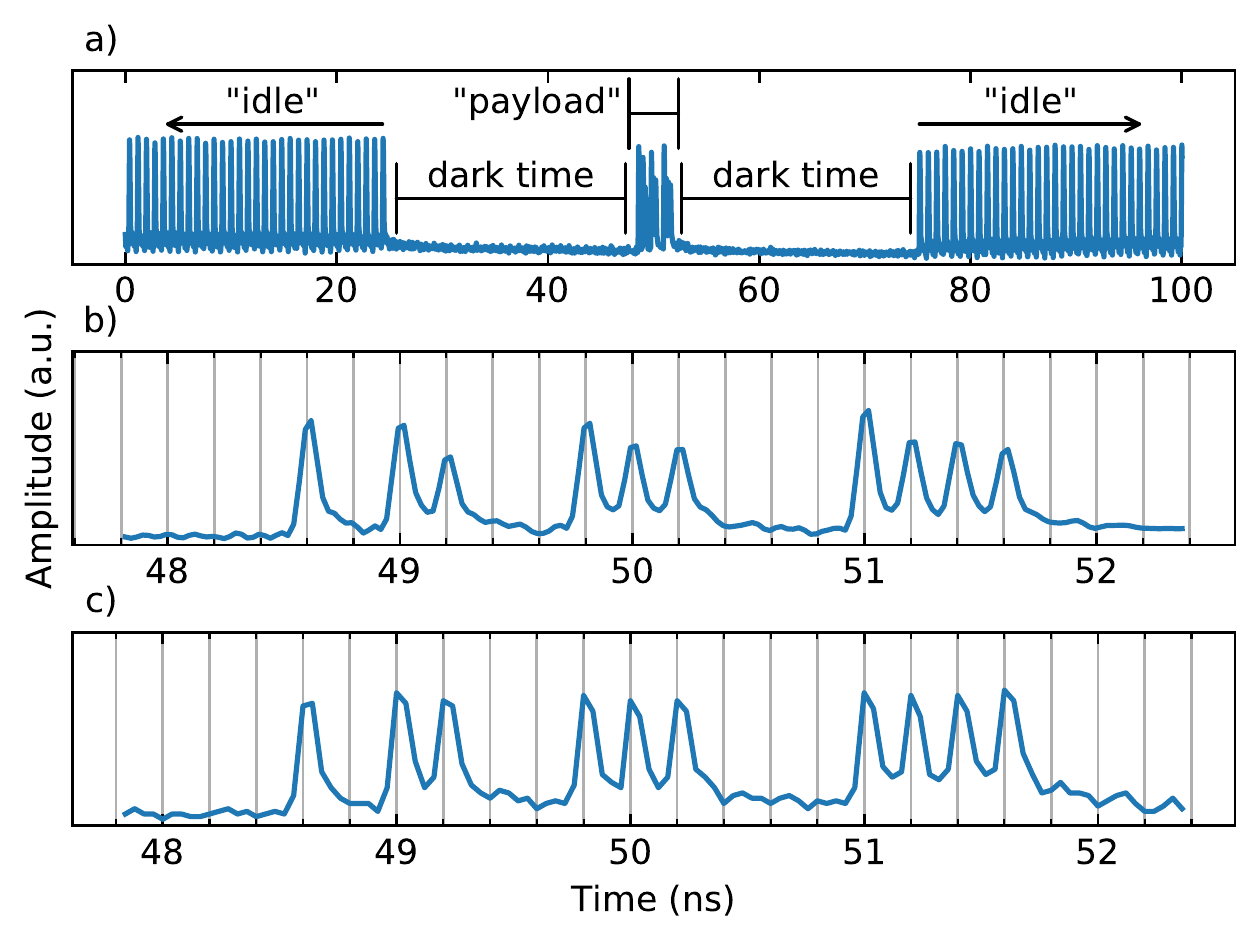}
\caption{Fast photodiode trace of residual 786 nm laser pulses exiting the PPKTP crystal.
(a) An arbitrary pulse pattern is created with the pulse picker at 1572 nm wavelength.
(b) Picking only the payload by using both pulse picker and Pockels cell.
Here, the BOA amplifies the pulses after the pulse picker to feed the forthcoming stage in the laser setup.
(c) Same as (b), but the BOA is replaced by an EDFA.
Now, all transmitted pulses have the same amplitude.
%shows a comparison and improvement in amplification of patterned pulses with a doped fiber amplifier.
Every grid line in the insets of (b) and (c) shows the location of a pulse in the 5 GHz continuous pulse train.
Figures (a) and (b) are adapted from \cite{Heinrich2019}.}
\label{switching}
%\end{center}
\end{figure}

\begin{figure}
\includegraphics[width=9cm,height=5cm,keepaspectratio]{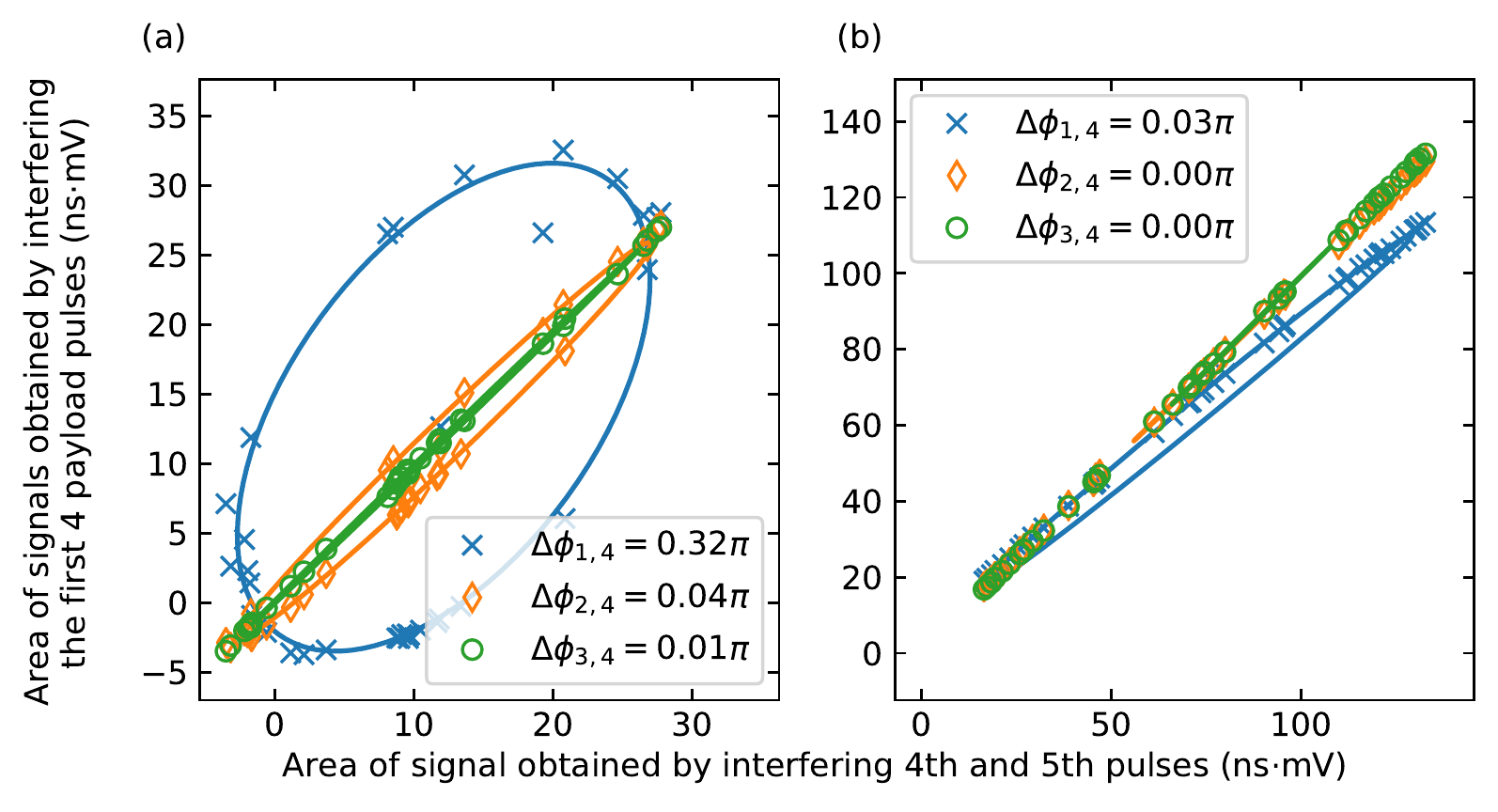}
\caption{Fitting fast photodiode data to measure relative phase shifts in the first five consecutive payload pulses at 5 GHz.
These 5 pulses are sent to a Michelson interferometer such that each pulse interferes with its predecessor with a random phase that is the same for all interfering pulse pairs (for details, see main text).
We measure the subsequent four interferometric pulses with a fast photodiode and extract the measured pulse area of each with an oscilloscope.
In the figure, the area of the first 3 interferometric pulses (blue crosses, yellow diamonds, green circles, respectively) is plotted against the area of the 4th pulse.
% %*************************************
% \textcolor{red}{By changing the optical path length difference of the interferometer on the order of $\delta$, where $\lambda \lesssim \delta \ll c \cdot \tau$, i.e., only small changes on the order of the wavelength and much less than the spatial pulse separation, we obtain data points which lie on an ellipse or a line segment.}
% %**********************************
(a) Fitting an ellipse to our data allows us to extract the phase shifts between the 1st \& 2nd, 2nd \& 3rd, and 3rd \& 4th pulses relative to the phase shift between the 4th \& 5th pulses. This data was taken when a BOA after the pulse picker was a part of the (old) setup.
%(b) Data is taken same as before but here BOA is replaced with EDFA, straight lines are hard to fit with an ellipse.
(b) Same as (a), but the BOA has been replaced with an EDFA.}
\label{ellipse}
%\end{center}
\end{figure}

To convert our pulsed laser into a fast entangling gate source, the device has to be capable of making pulse patterns by switching individual laser pulses.
For its realization, we employ two switching stages synchronized with a 250 MHz optical clock extracted directly from the frequency comb as shown in Fig. \ref{setup}.
Single-pulse switching is done in the first stage inside a pulse picker (Photline), constituted of a Mach-Zehnder interferometer and an electro-optic-modulator with 7 GHz bandwidth
that is driven by an arbitrary waveform generator with a data rate of 25 GS/s.
The fast pulse picker creates a \textit{payload signal} by picking individual pulses at 1572 nm as shown in Fig. \ref{switching}a.
In order to seed subsequent amplifiers, the pulse picker transmits a continuous pulse train at all other times, the \textit{idle signal}.
The second stage is a barium borate Pockels cell that serves as a slow switch with a minimum allowed switching time of 35 ns to filter the payload pulses by blocking the idle signal.
Previously \cite{Heinrich2019}, we observed that a pulse succeeding a blocked pulse---i.e., a dark time---had a different phase and amplitude with respect to a pulse succeeding a transmitted pulse as shown in Fig. \ref{switching}b.
We found that the BOA (which was installed right after the pulse picker) was responsible for these anomalies because the carrier injection and recombination time of such amplifiers is similar to the 200 ps pulse period \cite{bonk2013linear} of the pulse train.
The effective gain recovery time $\tau_g$ of this BOA was measured to be $\leq$500 ps.
This is the reason why the pulse anomalies were prominent at 5 GHz repetition rate (see Fig. \ref{ellipse}a) and disappeared at 1.25 GHz. The pulse repetition rate is changed by picking individual pulses inside the pulse picker.  

To overcome the aforementioned problem, a seemingly straightforward solution is to establish a sufficiently fast gain recovery ($\tau_g \ll$ 200 ps) during amplification. However, we could not find an easy solution for achieving such amplification at 5 GHz repetition rate that was compatible with our local pre-amplification laser parameters. On the other hand, we find that integrating an amplifier with a much slower gain recovery time i.e., $\tau_g \gg$ 200 ps can make the amplification immune to the fast changes due to a high repetition rate. Therefore, we substituted the BOA with a suitable EDFA with $\tau_g$ on the order of microseconds.
%To overcome this problem, we could not find an easy solution to establish a fast gain recovery ($\tau_g \ll$ 200 ps) for a pattern-effect-free amplification compatible with our laser parameters.
%Nonetheless, we decide to replace the BOA with a suitable EDFA because typically in doped fiber amplifiers $\tau_g \gg$ 200 ps, and therefore much longer than our pulse period, hence making the amplification immune to fast signal modulation.

The potential pulse-to-pulse phase shifts are verified and quantified by interfering the
pulses in a Michelson interferometer \cite{Heinrich}.
The beam splitter divides the payload pulses into two beams.
One of the beams is
time-delayed by one pulse period $\tau_\text{pulse}$ with respect to the other one before the beams are recombined on the interferometer's beam splitter.
In this way, every pulse $i$ interferes with the
successive pulse $i + 1$ in the payload.
The interference signals are recorded by a fast photodiode.
% when amplified with the BOA and its substitute (EDFA) separately.
The net phase difference $\Delta\Phi_{i}$ of the two
pulses at the output of the interferometer is a function of the phase
difference $\Delta\phi_{i}$ of the interfering pulses, and the path length difference
$\Delta x = 2(x_2 - x_1) = c \cdot \tau_\text{pulse}$ and
$\Delta\Phi_{i} = \Delta\phi_{i} + k \cdot \Delta x$,
where $x_1$ and $x_2$ are the lengths of interferometer arms, $c$ is the speed of light and $k$ the magnitude of the wave vector. The retro-reflecting mirror mounted on a manual translation stage is used to change $\Delta x$, therefore, the deterministic tuning of $\Delta x$ on a sub-wavelength scale is not possible. However, we find that over a time scale of about 100 ms, $\Delta x$ can be considered constant on a sub-wavelength scale.
Each measurement is repeated for a randomly chosen $\Delta x$.
By changing the optical path length difference of the interferometer on the order of $\delta$, where $\lambda \lesssim \delta \ll c \cdot \tau$, i.e., only small changes on the order of the wavelength and much less than the spatial pulse separation, we obtain data points which lie on an ellipse or a line segment, as shown in the Fig. \ref{ellipse}.

\section{UV pulse generation}
\label{UV}
In contrast to all previous stages (see the Fig. \ref{setup} panels A and B), high peak intensities are required for non-linear frequency conversion in order to maximize the output power.
However, given a constant average power, an increase in the repetition rate reduces peak intensity, which deteriorates the conversion process and forces us to make a trade-off between repetition rate and peak intensity.
To this end, we
(a) maximize the amplification as much as available (10 W EDFA) with all desired laser parameters at 1572 nm,
(b) minimize losses during post-amplification dechirping, and
(c) efficiently remove dispersion to attain transform-limited pulses.

\subsection{Dispersion control}
In order to satisfy (a), dispersion management in the laser system is crucial for achieving amplification that is free of self-phase-modulation.
We estimate the  dispersion required  to avoid inducing non-linearity in the high power EDFA to be $|D_\lambda| \sim12.5$ ps/nm.
% In order to satisfy (c) we need to carefully balance anomalous dispersion added by every meter of fiber in the system.
We optimize dispersion by adding and removing chirp in three stages:
First, we add a 
%TeraXion made 
chirped fiber Bragg grating (CFBG, made by TeraXion) 
just before the high power amplifier.
The CFBG provides a dispersion of -9.5 ps/nm.
Next, a chirped volume Bragg grating (CVBG, produced by OptiGrate), which adds +12.5 ps/nm of dispersion, is set up in the free-space output of the 10 W EDFA.
In the last step, we carefully balance the dispersion by adjusting the length of a stretcher fiber (inverse dispersion fiber with a dispersion parameter of -41 fs/nm/m at 1560 nm) in order to satisfy (c).
An optimum length of 100 m is sufficient to cancel the dispersion offered by the CVBG and the rest of the components before frequency up-conversion.
Prior to high-power amplification, pulses are passed through a circulator which first directs the laser pulses towards the CFBG and then to the 10 W EDFA.

Management of the entire dispersion in the stretcher fiber is not possible due to the short pulse period of 200 ps which is similar to the pulse length of the stretched pulses.
As a result, the pulse wings would overlap and we would forfeit the ability to separate the individual pulses for a clean pulse switching.
Therefore, it is critical to distribute chirping in the system.
The amplified (9.6 W average power) pulses are collimated and steered into the CVBG which reflects more than 90\% of light.
Temperature tuning of the CFBG enables us to precisely tune the dispersion by about $\pm 0.005$ ps/nm to achieve compressed 560 fs pulses.
The time-bandwidth product is measured to be 0.49, fairly close to a transform-limited pulse with a Gaussian pulse profile.
The efficient dechirping of stretched pulses is crucial to maximize peak intensities and thereby enhancing the efficiency of nonlinear frequency up-conversion.
We get 90\% efficiency in the compressor and hence 8.6 W average power is available for frequency up-conversion out of the pulse compressor.

\subsection{Pump pulse energy optimization}
\label{pumpopt}
\begin{figure}
\includegraphics[width=9cm,height=5cm,keepaspectratio]{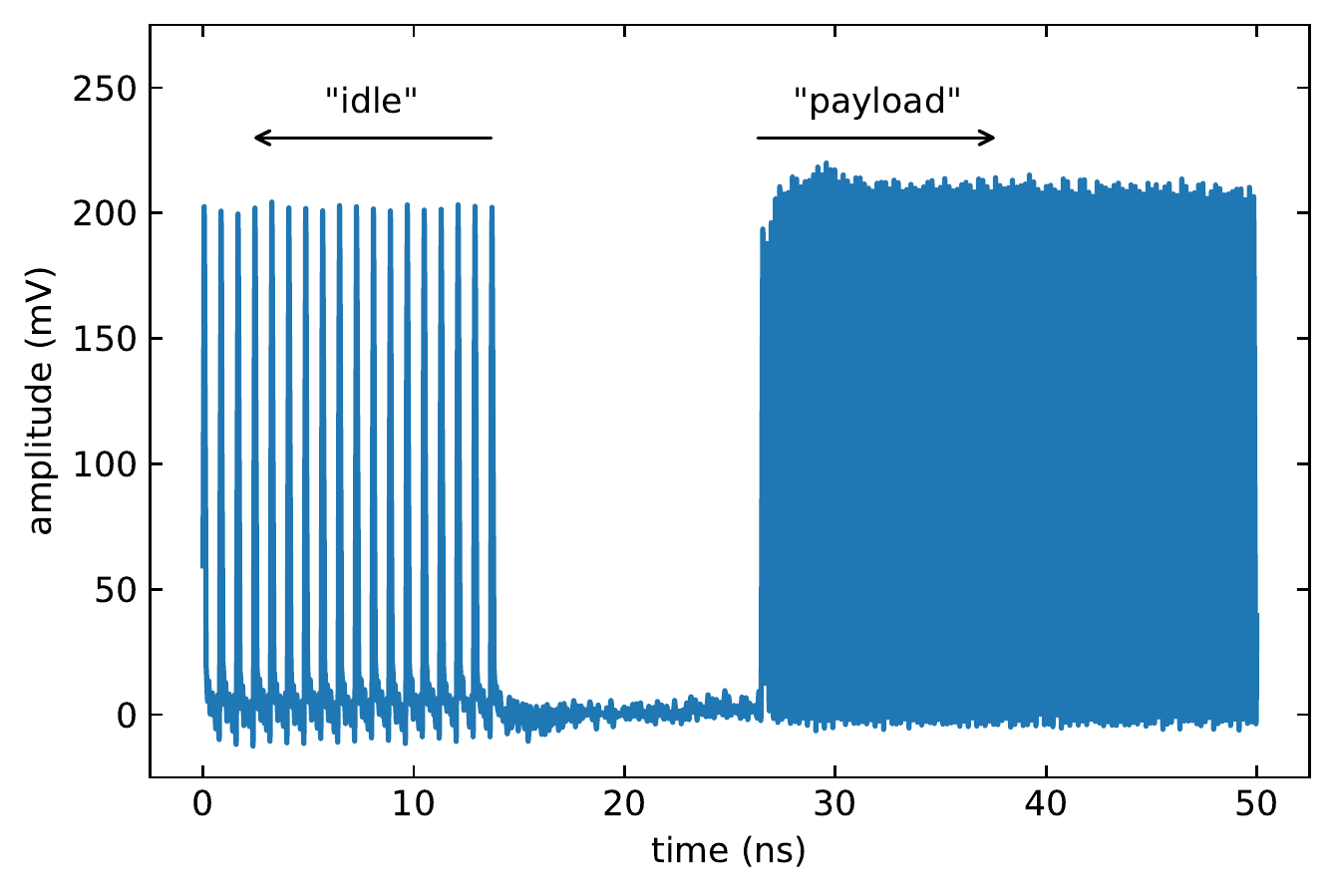}
\caption{Fast photodiode trace of laser pulses taken before the PPLN crystal, where the idle pulses have a 1.25 GHz repetition rate and the payload pulses have 5 GHz.
The change in the repetition rate changes the steady state of the 10 W amplifier,
i.e., it changes the pulse energy, but not the average power.
Before this change starts to happen, the pulse amplitude remains the same for at least 150 ns (only partially shown), or for 750 consecutive pulses at 5 GHz.
During this time and in this case, the average power is increased by a factor of 4 as compared to idle pulses.}
\label{timecons}
%\end{center}
\end{figure}

Our goal is to at least double the UV pulse energy compared to the previously available $\sim$ 80 pJ \cite{Heinrich2019}.
With the given fundamental pump power of 8.6 W, we did not succeed to significantly increase the pulse energy ($\geq$ 160 pJ) without lowering the repetition rate. Hence, we block three out of every four pulses, thereby dropping the repetition rate of the idle signal by a factor of four. The amplified average power remains the same at the output of 10 W EDFA at a lower repetition rate, therefore, the pump pulse energy is increased by the same factor of four.
We stop here because a further drop (\textless 1.25 GHz) in the repetition rate demands additional dispersion to avoid nonlinear pulse distortion during chirped pulse amplification. As the speed of the gate operation can be increased by ramping up the repetition rate, we keep the payload repetition rate at 5 GHz. %However, this will require extra control to alter the direction of momentum kicks.}
% \textcolor{red}{However, to satisfy the pulse energy
% requirement at 5 GHz repetition rate, we artificially retain the enhanced pulse energy in the payload pulses due
% to a relatively long time (hundreds of ns), which the 10
% W amplifier takes to fully settle to a new steady state
% after a change in the repetition rate, shown in Fig. 5.
% We intentionally place the payload pulses at the start of this new steady state transition.}
Using this approach, we satisfy the pulse energy requirement at 5 GHz repetition rate by retaining the enhanced pulse energy in the payload pulses. The key point here is to exploit the time, which the 10 W amplifier takes to fully settle into a new steady state after a change in the repetition rate of the idle and payload pulses prior to high power amplification shown in Fig. \ref{timecons}. The transition time for the amplifier to fully acquire a new steady state is hundreds of nanoseconds, which is long as compared to the duration of the payload pulses (plus the dark time between the end of the idle and the start of the payload period). This allows us to write a pulse picking/switching sequence such that we can intentionally place several hundreds of payload pulses well before the start of the new steady state transition.
As a consequence, the payload pulses contain the same pulse energy as in the idle pulses while keeping the maximum and unaltered repetition rate.
%We stop here because a further drop in the repetition rate would make the pulse-chirping insufficient to suppress the amplification induced non-linearity.
%Furthermore, by exploiting the long time-constant of the 10 W EDFA, we artificially retain the enhanced pulse energy without lowering the repetition rate in the payload, shown in Fig. \ref{timecons}.

\begin{figure}
\includegraphics[width=7.5cm,height=16cm,keepaspectratio]{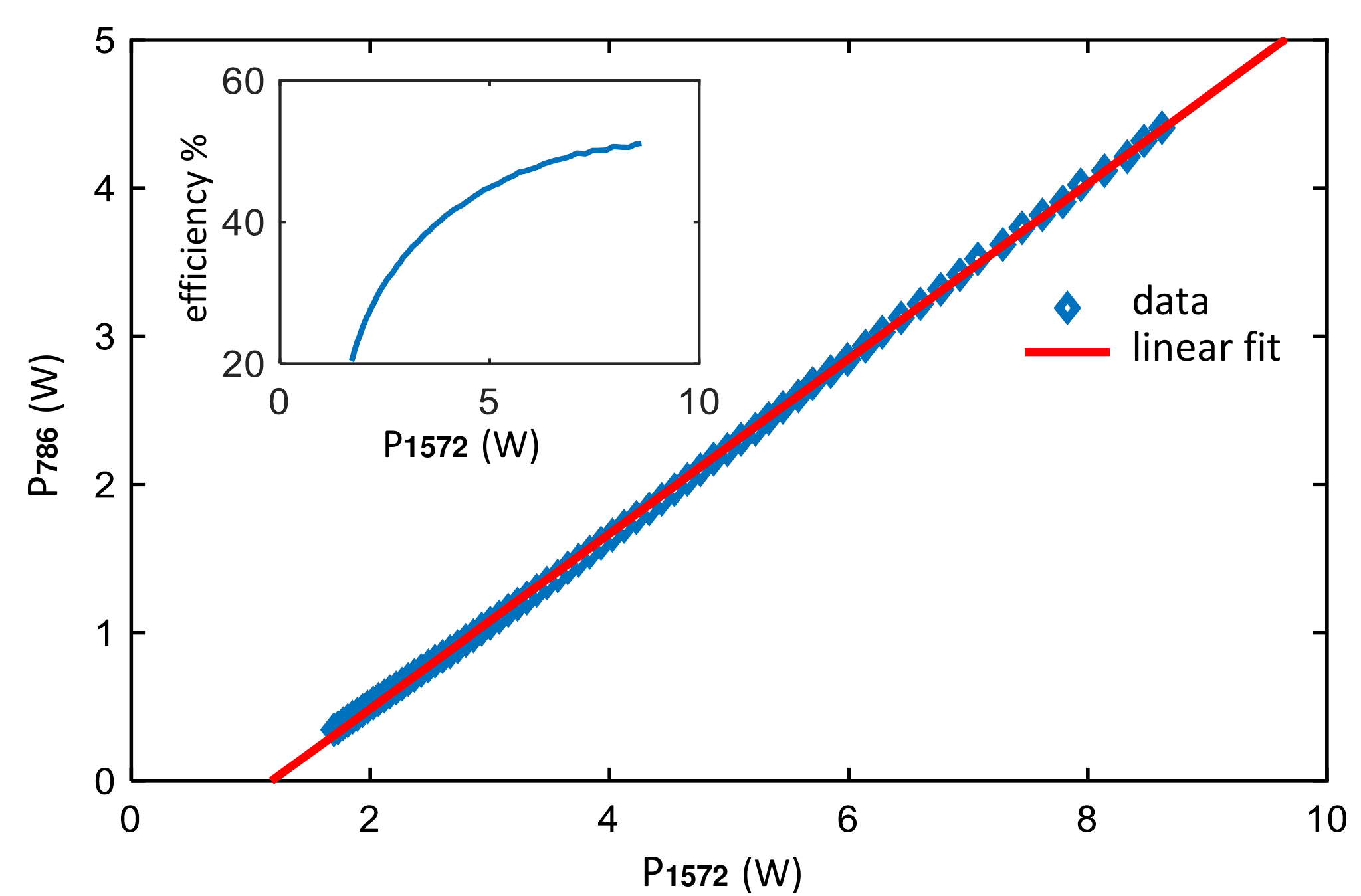}
\caption{Generated time-averaged power $P_{786}$ in the single-pass frequency conversion of the first doubling stage (1572 nm to 786 nm) as a function of the time-averaged pump power $P_{1572}$.
The inset shows the frequency conversion efficiency $P_{786}/P_{1572}$ with respect to the pump power.}
\label{conversion}
%\end{center}
\end{figure}

Under these conditions, vertically polarized  7 nJ laser pulses are focused down to 11 $\mu$m ($1/e^2$ spot radius) inside the periodically poled lithium niobate
(PPLN) crystal to achieve a maximum peak intensity of $\sim$ 3 GW/cm$^2$ at a given spot radius of a temporally Gaussian pulse profile.
The crystal is mounted inside a thermoelectric oven to achieve phase matching.
As a result, we get over 4.3 W of average power at 786 nm which corresponds to a pulse energy of 3.5 nJ and a single-pass second harmonic generation (SHG) efficiency of 50\% as shown in the inset of Fig. \ref{conversion}.
The SHG efficiency curve is not linear because the second harmonic efficiency starts saturating when the pump power exceeds approximately 2.5 W. This saturation can be attributed to the pump depletion and thermal phase-mismatch in the PPLN crystal for a given focusing condition.

\subsection{UV pulse energy estimation}
\label{UVenergy}

%[[ The entangling gate requires only a few UV pulses at a high repetition rate and a sufficiently high average power. Therefore, we have to make a trade-off and favor a higher non-linear coefficient over a wide UV transparency range
%and choose non-critically phase-matched periodically poled potassium titanyl phosphate (PPKTP) instead of a conventional borate family crystal to generate 393 nm pulses.]]

The entangling gate operation requires only a few UV pulses at a high repetition rate which allows us to overcome thermal effects due to absorption inside the non-linear crystal.
Hence, we can make a trade-off and favor a higher non-linear coefficient over a wide UV transparency range
and choose non-critically phase-matched periodically poled potassium titanyl phosphate (PPKTP) instead of a conventional borate family crystal to generate 393 nm pulses.

For a pump power of 500 mW at 786 nm and 1.25 GHz repetition rate we get over 100 mW of 393 nm light focused to a spot size with a $1/e^2$ radius of 8.5 $\mu$m. Cross-correlation frequency-resolved optical gating measurement gives a UV pulse duration on the order of a picosecond.
At higher pump intensities thermal-dephasing induced power fluctuations in the  PPKTP crystal become problematic, as reported in earlier studies \cite{lundeman2009threshold,han2014generation}.
As a consequence, it is difficult to gauge the exact power at the crystal output.
Therefore, at a higher pump power, the duty cycle $D_C$ of the pump beam has to be reduced to get a stable output-power reading for pulse energy estimation.
We choose a period $T = 1$ ms and different pulse widths $PW$ which yield $D_C = PW / T$.
We use three different ways to measure the pulse energy. The measurements agree with each other with deviations of a few percent.
%First, we determine the duty cycle which allows us to get stable UV power at maximum pump power and estimate UV power $\geq$ 1 W.
%Here, one complete cycle is 1 ms long where the pulses are ON for a 1 us.
First, for several different $D_C \leq 1/1000$, we measure the average UV power at maximum pump power and estimate a UV power $\geq$ 1 W.
We verify the method by measuring (1) the minimum pump power, (2) the maximum pump power, (3) the UV power at minimum pump power, each with and without duty cycle---i.e., $D_C \ll 1$ and $D_C = 1$, respectively---%
%We use the same duty cycle to calibrate power meter by monitoring pump power with and without duty cycle
and find an excellent agreement between actual and projected power numbers.
Second, we use only a payload which comprises 50 pulses every 1 ms and get  $\sim$ 50 $\mu$W average power, which corresponds to  $\sim$ 5 W of average power for a continuous pulse train at 5 GHz.
The reason behind this huge power is a sudden increase in the repetition rate (5 GHz) of payload pulses while the amplifier is being operated in the steady state of 1.25 GHz.
This also corroborates that the UV power with a continuous pulse train at 1.25 GHz is over 1 W.
Third, we monitor the payload signal with a fast photodiode and oscilloscope to further verify the change in the pulse amplitude with the pump power.
We find that the amplitude of the peak photodiode voltage grows steadily from 4 mV (100 mW UV power) to a maximum of 45 mV, which thus also corresponds to a UV power of $\geq$ 1 W.
The fractional peak-to-peak amplitude noise is below 5\%. 
For the given UV average power we determine a pulse energy $\geq$ 800 pJ even for a payload repetition rate of 5 GHz.

\section{Fast gate simulations}
\label{simulation}
\begin{figure}
\includegraphics[width=7.5cm,height=12.5cm,keepaspectratio]{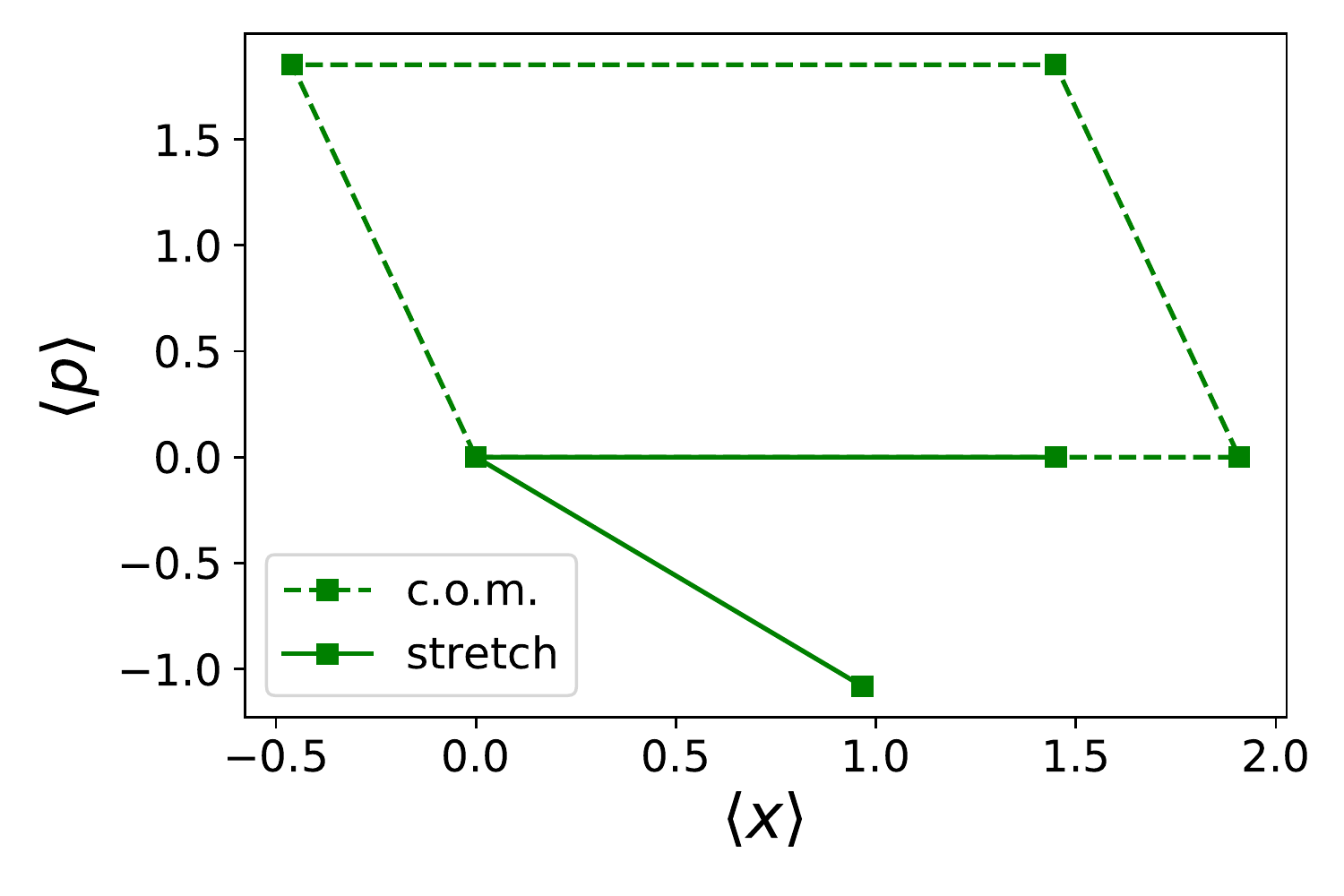}
\caption{Phase-space trajectory of the center of mass (dashed) and stretching (solid) modes in the frame of reference that rotates with the frequency of each mode $\braket{a e^{i\omega_{c,s}t}}=(\braket{x_{c,s}}+i\braket{p_{c,s}})/\sqrt{2}$ for a gate implemented by a sequence with 4 pulses. 
Only the center-of-mass motion contributes to the total phase $\phi$ in this particular gate.}
\label{gate}
\end{figure}
The experimental setup described in \cite{Heinrich2019} combined with the ultrafast pulsed laser source based on a frequency comb presented above constitute the basis for the implementation of a fast two-qubit phase gate \cite{Torrontegui2020}. We propose an implementation using $^{40}\mbox{Ca}^+$ ions where the qubit is stored in the $4\mbox{S}_{1/2}$ and $3\mbox{D}_{5/2}$ internal states and we use the $4\mbox{S}_{1/2}\leftrightarrow 4\mbox{P}_{3/2}$ transition for applying state-dependent momentum kicks to the ions via resonant laser pulses. Two trapped ions loaded in a common 1D harmonic potential of frequency $\omega$ freely evolve under
the Hamiltonian $H_0=\hbar\omega_ca_c^{\dag}a_c+\hbar\omega_sa_s^{\dag}a_s$ with $\omega_c=\omega$
and $\omega_s=\omega\sqrt{3}$ and $a_{c,s}^{\dag}$ $(a_{c,s})$ the creation (annihilation) phonon operator for the center-of-mass and stretch modes. This free evolution is interleaved with a fast pulsed interaction with 
an on-resonant laser beam kicking the ion described by $H_1=\Omega(t)[\sigma_1^{\dag}e^{i\hbar k x_1}+\sigma_2^{\dag}e^{i\hbar k x_2}+\mbox{H.c.}]/2$ 
with $\sigma_i^{\dag}$ the ladder spin operator for the $i$-th ion \cite{cirac1995quantum,garcia2003speed, Garcia-Ripoll2005}, and $\Omega(t)$ is the Rabi frequency.
The pulse duration $\delta t$ and the pulse period are orders of magnitude faster than the trap period to fully excite and quickly de-excite the population via stimulated emission, i.e. $\int_{0}^{\delta t}\Omega(t)dt=\pi$. Additionally, each pulse is split by a 50/50 beam splitter such that the resulting two pulses counter-propagate arriving at the ion with a relative delay $t_{wait}$, controlled by the relative length of the two optical paths. In this manner, both the length of the pulses $\delta t$ and the delay $t_{wait}$ between counter-propagating pulses are shorter than the lifetime of the $4\mbox{P}_{3/2}$ state, $\delta t,t_{wait}\ll t_{\gamma}=6.9$ ns. As consequence of the $\hat H_1$ interaction the ion is excited by the first pulse acquiring a momentum $ \hbar k$, along the direction of $k$. The second pulse coming from the opposite direction coherently de-excites the ion, providing the same amount of momentum, but in the opposite direction of $k$. As discussed below, spontaneous emission during two consecutive pulses will define the ultimate fidelity that can be achieved in the resonant excitation scheme.  Finally, the repetition rate of the laser is considered to be much faster than the trap frequency, allowing a fine-grained control of the pulse sequences.

The unitary operation generated by a sequence containing $N$ pulses expressed in the center-of-mass and stretch coordinates takes the form $\mathcal{U}=\mathcal{U}_c\mathcal{U}_s$, where $\mathcal{U}_{c,s}=\prod_{n=1}^NU_{c,s}(t_n)$ are the unitary operators produced by $H_1$ and an interspersed free evolution $H_0$ during a $t_n$ time \cite{Torrontegui2020}. In phase space, the Heisenberg representation allows us to interpret these interactions as displacements of the operators 
$a_{c,s}$ by a complex number $A_{c,s}$ that depends on the collective state of the ions
\beqa
\label{disp}
a_c&\rightarrow&a_c+A_c=a_c+i(\sigma_1^z+\sigma_2^z)\alpha_c\sum_ne^{-i\omega_ct_n}, \nonumber \\
a_s&\rightarrow&a_s+A_s=a_s+i(\sigma_1^z-\sigma_2^z)\alpha_s\sum_ne^{-i\omega_st_n},
\eeqa
with strengths $\alpha_c=\eta/2^{3/2}$ and $\alpha_s=\alpha_c/3^{1/4}$
proportional to the Lamb-Dicke parameter $\eta=k\sqrt{\hbar/(2m\omega)}$. 
The normal modes follow polygonal orbits, see Fig. \ref{gate}, where the edges all have uniform length $\sim\alpha_{c,s}$ and controllability of the kicking sequence is limited to the allocation of the pulse arrival times $\omega t_n$ that determines the relative angle between edges. In case that both modes restore the initial position in phase-space, 
\beq
\label{conmensurability}
A_c=A_s=0, 
\eeq
the resulting unitary is equivalent to a free evolution up to a global phase, which is independent of the motional states
\beq
\label{phi}
\phi=\alpha_c^2\sum_{j=2}^N\sum_{k=1}^{j-1}\bigg[\frac{\sin(\sqrt{3}\omega t_{jk})}{\sqrt{3}}-\sin(\omega t_{jk})\bigg],
\eeq
with $t_{jk}=t_j-t_k$. When this phase becomes
\beq
\label{phase}
\phi=\pi/4+ 2n\pi \quad n=0,1,2\dots
\eeq 
the free evolution corresponds to a controlled-phase gate.

In the following we solve the set of equations (\ref{conmensurability}) and (\ref{phase}). 
 Due to the finite repetition rate of the source generator,
the pulse sequence design combines {\itshape i)} simple
continuous solutions with {\itshape ii)} fine-tuned pulse picking based on a genetic algorithm \cite{Torrontegui2020} leading to optimal solutions that minimize the gate error $\epsilon=|A_c|^2+|A_s|^2$ 
that quantifies the deviations to recover the original position after the whole kicking sequence. The fastest allowed gate will depend on the extent of control. However, this simple scenario which relies only on pulse picking \cite{Torrontegui2020}, 
and fixes the strength, direction, and repetition rate of the pulses, already leads to gates faster than the trapping period $T<2\pi/\omega$.

The fastest sequence fulfilling Eqs. (\ref{conmensurability}) and (\ref{phase}) corresponds to a sequence of 4 pulses and a trapping 
frequency of $\omega \sim 2\pi\times 0.27$ MHz. The resulting phase gate is implemented in a time $T\sim0.79 \times 2\pi/\omega$ (or $<$ 3 $\mu$s), assuming a repetition rate of $\sim 5$ GHz. The phase-space trajectories of the center-of-mass and stretching modes for such gate are depicted in Fig. \ref{gate}.
The gate time is expressed in terms of $2\pi/\omega$, the harmonic oscillator characteristic time-scale period that accounts for the phononic motion.
Theoretically, the proposed scheme using sequences of resonant pulses does not restrict the achievable acceleration of the two-qubit gates. For example, by only adding directionality, i.e., allowing different directions of the kick can lead to the design of pulse sequences that reduces the gate time when increasing the number of pulses. Additionally, if we provide more controllability, tilting the laser pulses with respect to the axial trapping axis \cite{garcia2003speed} allows for even faster gates. This extra control can be an experimental challenge and may induce other sources of errors to be analyzed in future works.

In the protocols discussed here, the motional state of the ion is almost perfectly restored and the gate fidelity is fundamentally limited by the errors in the internal state of the ion. More precisely, our method for kicking the ion implies that the atom spends some time in the excited state $4\mathrm{P}_{3/2}$. We can safely assume short pulses with a duration $\delta t \simeq 1$ ps and a spacing $t_{wait}\simeq 1$ ps much shorter than the inverse of the spontaneous emission rate $1/\gamma\simeq 7$ ns. Under these conditions, the probability that the ion relaxes, emitting a photon, is essentially $\epsilon_1 \sim \mathcal{O}(\gamma t_{wait})\sim 1.8\times 10^{-4}.$ If we have a sequence of $N$ kicks, the infidelity of the gate operation is approximately $\epsilon_N=1-(1-\epsilon)^N\sim \mathcal{O}(N t_{wait}\gamma)$ or $\epsilon_4 \simeq 7.4\times 10^{-4}$ for the optimal sequence described above.

This error can be decreased by using shorter pulses or substituting the resonant $4\mathrm{S}_{1/2}\leftrightarrow 4\mathrm{P}_{3/2}$ transition with a Raman or STIRAP process that connects the qubit states $4\mathrm{S}_{1/2}\to 3\mathrm{D}_{5/2}.$ The energy difference between S and D states is still sufficiently large such that a single transition will impart a significant momentum to the ion. However, the spontaneous decay rate of the $D$ state ($\sim$1.1 s) is orders of magnitude smaller, eliminating the fundamental limitation due to spontaneous emission.

Note that faster gates acting on a time $T\sim 0.25 \times 2\pi/\omega$ have been recently designed \cite{Gale2020}. However, this theoretical limit remains a technical challenge because it requires switching the directions of the laser pulses---i.e., choice of which pulse arrives first in the counter-propagating pair---, a process that may induce additional sources of error and deteriorate the gate fidelity \cite{mizrahi2013ultrafast}. Another approach based on pulse shaping techniques has produced optimal continuous protocols \cite{schafer2018fast} with fast experimental gates of around $T\sim 0.89 \times 2\pi/\omega$, but they still exhibit a rather poor fidelity of $0.6$.

\begin{figure}
\includegraphics[width=9cm,height=4.5cm]{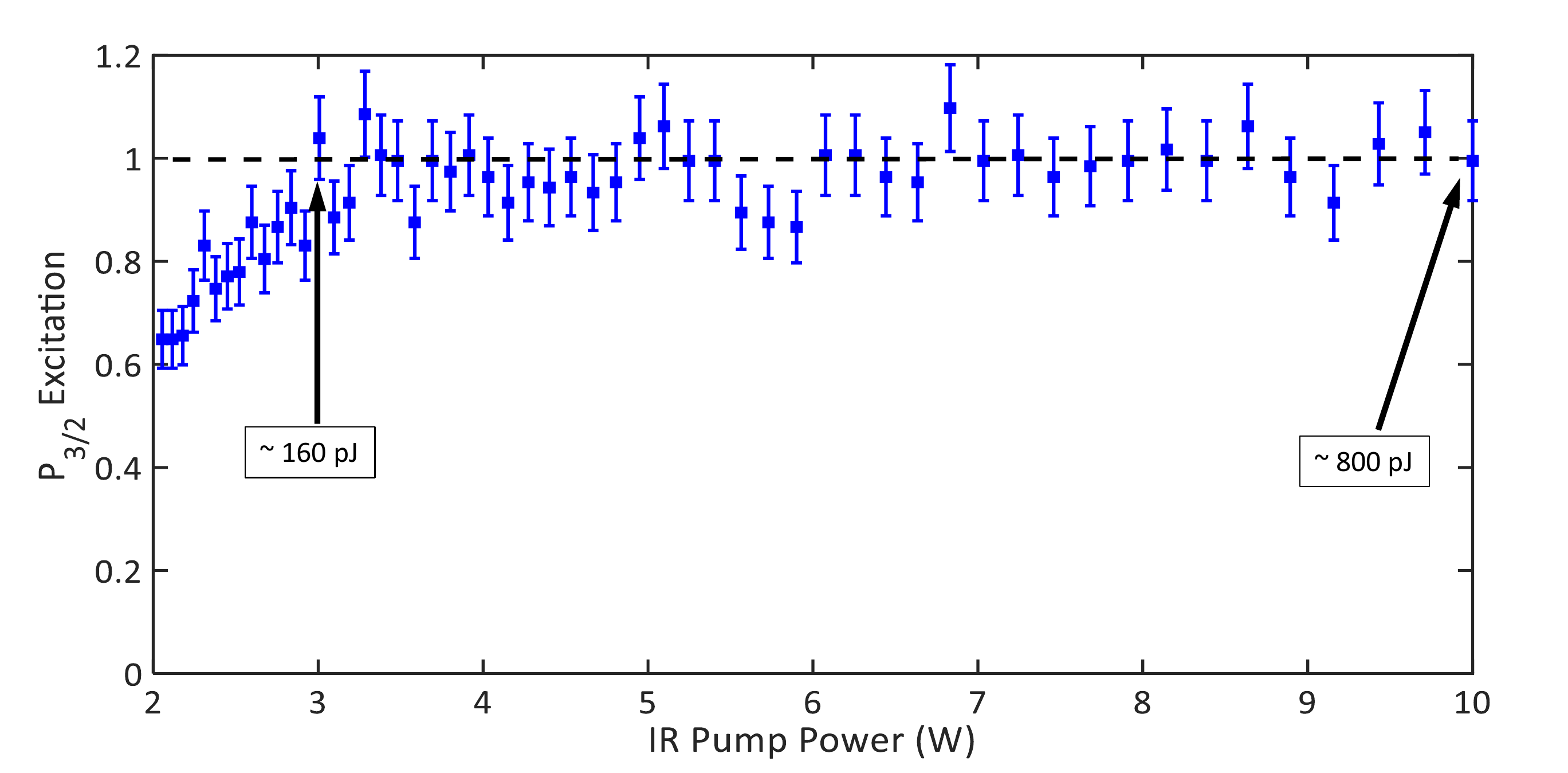}
\caption{Demonstration of a rapid adiabatic passage on the $\mbox{4S}_{1/2}\leftrightarrow\mbox{4P}_{3/2}$ transition in Ca$^+$. Each data point is taken by scanning the infrared pump power. Black arrows denote the corresponding UV pulse energy measured after PPKTP crystal that is high enough to transfer the population to the excited state with near-unit efficiency. The power tunability range is dependent on the minimum to maximum output power of the 10 W EDFA (2 W to 10 W). The dashed horizontal line marks the unit probability as a guide to the eye.}
\label{rap}
\end{figure}

\section{Trapped-ion rapid adiabatic passage \label{sec:TrappedIonExperiment}}

In order to characterize the generated 393 nm laser pulses, we subject a Ca$^+$ ion prepared in 4S$_{1/2}$ to a single ps-pulse and observe a rapid adiabatic passage (RAP) between the $4\mbox{S}_{1/2}\leftrightarrow 4\mbox{P}_{3/2}$ states.

A pulse shaper composed of a grating pair each with groove density 3040/mm is employed after the PPKTP crystal and before UV pulse delivery fiber (see Fig. \ref{setup}). This adds a theoretically estimated group delay dispersion of $>$ 5 ps$^{2}$ in the 393 nm pulses. Under optimum conditions, the pulse shaper efficiency is found to be $\leq$ 50 \% after a total of four passes through the grating pair. These temporally shaped pulses are coupled into the fiber and focused on the ion with a waist of $\sim 6\,\mu$m.

Fig. \ref{rap} shows the probability of exciting the ion to the 4P$_{3/2}$ state by a single pulse as a function of the pulse energy. As the excited state decays on a time scale of a few nanoseconds back to 4S$_{1/2}$ and to the metastable 3D-states, we use the well-known branching ratio \cite{Gerritsma:2008} for extracting the P$_{3/2}$ population by monitoring the fraction of events where the ion decays to the 3D$_{5/2}$ state \cite{Heinrich2019}. The data shows that for sufficiently high pulse intensities, the excitation probability $p$ is very close to unity over a wide range of pulse intensities. We did not try to constrain $p$ to the physically valid range of $[0,1]$, as data points with $p>1$ result just from the finite number of experimental repetitions and constraining them to $p\le 1$ would have introduced a systematical bias.

\section{Conclusion and Outlook}
\label{conclude}
Bursts of UV pulses at 5 GHz repetition rate with an average power of 5 W have been reported, while keeping the bandwidth small ($\sim$ 500 GHz).

The resulting pulse energy is $\sim$ 10 times higher than the previous demonstration for a coherent population transfer in a \Ca{} ion. This will enable us to create a pulse pair from a single pulse to generate  $2\hbar k$ momentum kicks for the implementation of fast entangling gate operations.
By adding a linear chirp to the UV pulses, we implement a RAP, which can make coherent excitations robust against pulse intensity fluctuations \cite{malinovsky2001general}. It is noteworthy that, although intensity fluctuations might give rise to phase fluctuations between the qubit states in a single RAP, in an entangling gate these RAPs come in pairs, which results in a cancellation of phase fluctuations provided that the intensity noise is slow compared to the gate duration. 

The enhanced pulse energy allows us to overcome the diffraction losses inside the pulse shaper while adding a required linear chirp in the pulses. It is desirable to find the optimum pulse shaper design where the necessary and sufficient dispersion conditions can be fulfilled to attain minimum possible delay in the counter-propagating pulse pair to avoid spontaneous emission.
\begin{acknowledgements}
We acknowledge support from Project PGC2018-094792-B-I00 (MCIU/AEI/FEDER,UE), CSIC Research Platform PTI-001, and CAM/FEDER Project No. S2018/TCS-4342 (QUITEMAD-CM). Additionally, we acknowledge funding by the Institut f\"ur Quanteninformation GmbH.\end{acknowledgements}
\providecommand{\noopsort}[1]{}\providecommand{\singleletter}[1]{#1}%

%\bibliography{mybib.bib}

\end{document}